\documentclass[twocolumn,aps,pra,superscriptaddress,showpacs,tightenlines]{revtex4}
\usepackage[T1]{fontenc}
\usepackage[latin9]{inputenc}
\usepackage{amsmath}
\usepackage{graphicx}
\usepackage{amssymb}

\begin{document}

\title{On quantum optical properties of single-walled carbon nanotube}

\author{Z. L. Guo}

\affiliation{School of Physics, Peking University, Beijing, 100871,
China}

\author{Z. R. Gong}

\affiliation{Institute of Theoretical Physics, The Chinese Academy of Sciences,
Beijing, 100080, China}

\author{C. P. Sun}

\affiliation{Institute of Theoretical Physics, The Chinese Academy of Sciences,
Beijing, 100080, China}

\date{\today}
\begin{abstract}
We study quantum optical properties of the single-walled carbon
nanotube (SWCNT) by introducing the effective interaction between
the quantized electromagnetic field and the confined electrons in
the SWCNT. Our purpose is to explore the quantum natures of electron
transport in the SWCNT by probing its various quantum optical
properties relevant to quantum coherence, such as the interference
of the scattered and emitted photons, and the bunching and
anti-bunching of photons which are characterized by the higher order
coherence functions. In the strong field limit, we study the
interband Rabi oscillation of electrons driven by a classical light.
We also investigate the possible lasing mechanism in superradiation
of coherent electrons in a SWCNT driven by a light pump or electron
injection, which generate electron population inversion in the
higher energy-band of SWCNT.
\end{abstract}

\pacs{78.67.Ch, 78.55.-m, 81.07.-b}

\maketitle

\section{\label{sec:introduction}INTRODUCTION}

Carbon nanotubes (CNTs) have been under great focus these years
because of their promising thermal and electrical conductivities,
and other unusual features that may lead to new
applications~\cite{Carbon1,Carbon2,Carbon3}. In recent years,
individual single-walled carbon nanotubes (SWCNTs) have
experimentally become available for the design of future quantum
devices~\cite{SWCNT1,SWCNT2,SWCNT3,SWCNT4}. Through putting such a
SWCNT between electrodes while maintaining a low contact resistance,
novel CMOS devices can be made from this novel
material~\cite{field1,field2,field3}. Surpassing the current
silicon-based CMOS devices, CNT-based CMOS devices appear to have
the potential for wide applications. To this end, a broad research
is required on various aspects of its characteristics beforehand.

The conventional investigation for a new material is to explore its
photoluminescence~\cite{optical1,optical2,optical3,optical4,optical5,optical6,optical7,optical8,optical9}
. We usually study the characteristic spectroscopy of the light
scattered by or emitted from this material. Meanwhile, since
ballistic transport--a motion of electrons with negligible
electrical resistivity due to scattering in the process of
transportation--happens in a SWCNT at low
temperature~\cite{field1,ballistic}, SWCNT should be treated beyond
the classical scenario, and pure quantum effects should be taken
into account. As a result, not only should the classical optical
properties (e.g., the intensity, the spectrum, etc,) of the SWCNT be
considered, but also the quantum optical properties (e.g., the
bunching and antibunching phenomena, etc,) need to be studied in
details. In this paper we develop a fully quantum approach for the
SWCNT-light interaction to address the quantum effects relevant to
the higher order quantum coherence. Our investigation is oriented by
the great potential to implement the quantum optical devices based
on current carbon nanotube technology, which works in the quantum
regime, or at a level of single quantum state.

Starting from the minimal coupling theorem, we derive the effective
Hamiltonian of the SWCNT interacting with a fully quantized light
field. The interband Rabi oscillation is first studied for the light
field whose intensity is sufficiently strong to be treated
classically. We explore the full quantum features of the
transporting electron in the SWCNT which is displayed by its quantum
optical properties. To this end, we quantize the light field
interacting with the confined electrons in SWCNT, and calculate and
analyze the higher order coherence functions of the photons
scattered or emitted from the SWCNT. It is shown that the total
population inversion of electrons, the first order and the second
order coherence functions strongly depends on the chiral vector of
the SWCNT, while this dependence does not exist in the generic
graphene. Additionally, the anti-bunching feature of the light field
is predicted with detailed calculations based on the long time
approximation. A similar discovery has been made in an
experiment~\cite{ballistic}, but to our best knowledge no
microscopic theoretical explanation has been given.

This paper is organized as follows. In Sec.~\ref{sec:setup}, the
interaction between the quantized light field and the SWCNT based
on the tight binding approach is derived from the the minimal coupling
theorem. In Sec.~\ref{sec:Interband}, we study the interband Rabi
oscillation of the electrons in the SWCNT induced by strong light
when the driving light can be treated classically, the reason of which
is generally proved in App.~\ref{App:strong}. The interference of
the scattered light from the SWCNT and the second order correlation
of the emitted photons are investigated in Sec.~\ref{sec:interference}
and Sec.~\ref{sec:Second}, respectively. Additionally, the possible
lasing mechanism of the SWCNT through a light pump or electron injection
is discussed in Sec.~\ref{sec:Possible}. The conclusions are presented
in Sec.~\ref{sec:conclusion}.

\section{\label{sec:setup}MODEL SETUP}

\begin{figure}[ptb]

\begin{centering}
\includegraphics[bb=44 95 527 700,clip,width=2.5in]{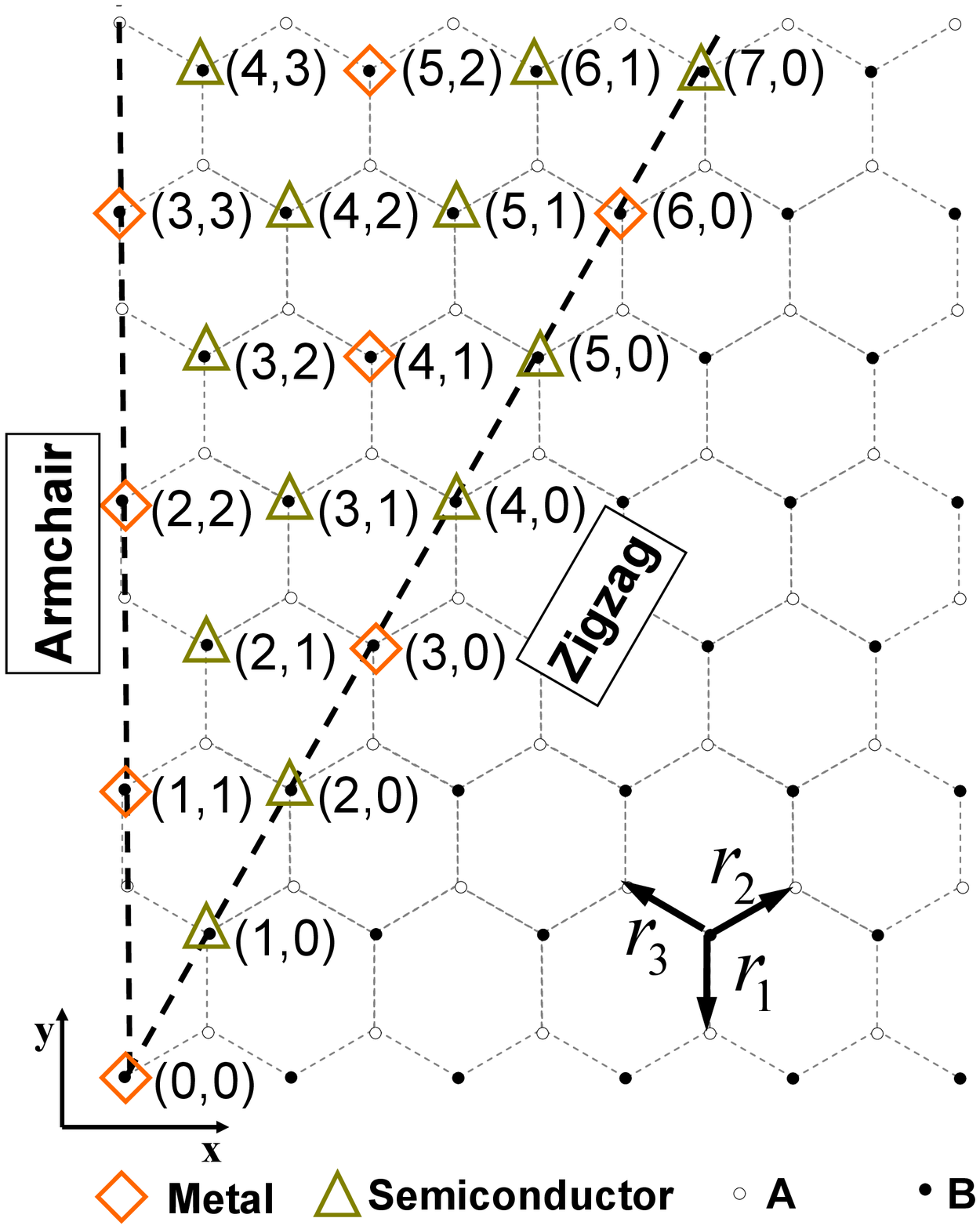}
\par\end{centering}

\caption{Schematic illustration of the $2$-D hexagonal lattice of
the SWCNT, which contain two sets of sublattices A and B. The pair
numbers $(n,m)$ denotes the chiral vector. }

\label{fig:model}
\end{figure}


The difference between carbon nanotubes and graphene is that carbon
nanotubes allow merely discrete wave vectors along their specific
chiral vector while graphene allows continuous ones, as long as we
neglect such effects as distortion of the lattice in carbon
nanotubes. Thus, to simplify the modeling of the system in
consideration, we can take the tight banding model of graphene into
account, and then apply discrete wave vector restriction to
demonstrate the properties of the nanotube. The honeycomb lattice of
graphene is divided into two triangular sublattices $A$\ and $B$ \
(see Fig. \ref{fig:model}). Here, the chiral vector of the SWCNT is
denoted as a pair of numbers $(n,m).$ The discrete wave vectors for
carbon nanotubes will be directly introduced by boundary conditions
later.

Since electrons in graphene approximately hop from one site to the
nearest neighbor one, a tight binding model \begin{equation}
H_{e}=-J\sum\limits _{\mathbf{r}\in A}\sum_{\alpha=1}^{3}[a^{\dagger}(\mathbf{r})b(\mathbf{r}+\mathbf{r}_{\alpha})+h.c.]\label{2-1}\end{equation}
 is applied to describe the motion of the electrons in the graphene.
Here, $J$ is the hopping constant; $a\ (a^{\dagger})$ and $b\
(b^{\dagger})$ annihilates (creates) an electron at sublattice $A$
and $B$, respectively. And $\mathbf{r}_{\alpha}$ $(\alpha=1,2,3)$
are the real space vectors pointing from one site to its nearest
neighbors. Usually they are chosen as \begin{subequations}
\begin{eqnarray}
\mathbf{r}_{1} & = & \frac{l}{\sqrt{3}}(0,-1),\label{2-2-1}\\
\mathbf{r}_{2} & = & \frac{l}{\sqrt{3}}(\frac{\sqrt{3}}{2},\frac{1}{2}),\label{2-2-2}\\
\mathbf{r}_{3} & = & \frac{l}{\sqrt{3}}(-\frac{\sqrt{3}}{2},\frac{1}{2})\label{2-2-3}\end{eqnarray}
 , and are schematically plotted in Fig. \ref{fig:model}. Here, $l$
is the lattice constant of both the triangular sublattice $A$ and
$B.$

To diagonalize the above tight banding \ Hamiltonian, \ a 2D Fourier
transformation \end{subequations} \begin{equation}
c_{\mathbf{k}}=\sum\limits _{\mathbf{r}\in C}c(\mathbf{r})e^{-i\mathbf{k}\cdot\mathbf{r}},(c=a\text{ or }b,C=A\text{ or }B).\label{2-3}\end{equation}
 is used to give the momentum space- representation of the Hamiltonian
(\ref{2-1}) \begin{equation}
H_{e}=\sum\limits _{\mathbf{k}}\left(\Phi_{\mathbf{k}}a_{\mathbf{k}}^{\dagger}b_{\mathbf{k}}+h.c.\right).\label{2-4}\end{equation}
 Here the transition energy $\Phi_{\mathbf{k}}\equiv-J\sum\limits _{\mathbf{\delta}\in\{\mathbf{r}_{\alpha}\}}e^{i\mathbf{k}\cdot\mathbf{\delta}}$
is a summation over all the directions of nearest neighbors. It is
explicitly written as \begin{equation}
\Phi_{\mathbf{k}}=-Je^{i\frac{k_{x}l}{\sqrt{3}}}\left(1+2\cos\frac{k_{y}l}{2}e^{-i\frac{\sqrt{3}k_{x}l}{2}}\right).\label{2-5}\end{equation}
 and corresponds to the transition of electrons between two sublattices
$A$ and $B$. Further, this Hamiltonian (\ref{2-4}) is diagonalized
as\begin{equation}
H_{e}=\sum\limits _{\mathbf{k}}E_{\mathbf{k}}\left(\alpha_{\mathbf{k}}^{\dagger}\alpha_{\mathbf{k}}-\beta_{\mathbf{k}}^{\dagger}\beta_{\mathbf{k}}\right)\label{2-6}\end{equation}
 through a unitary transformation \begin{subequations} \begin{eqnarray}
\alpha_{\mathbf{k}} & = & \frac{1}{\sqrt{2}}\left(e^{-i\varphi_{\mathbf{k}}}a_{\mathbf{k}}+e^{i\varphi_{\mathbf{k}}}b_{\mathbf{k}}\right),\label{2-7-1}\\
\beta_{\mathbf{k}} & = & \frac{1}{\sqrt{2}}\left(e^{-i\varphi_{\mathbf{k}}}a_{\mathbf{k}}-e^{i\varphi_{\mathbf{k}}}b_{\mathbf{k}}\right).\label{2-7-2}\end{eqnarray}
 Here, the single particle spectrum is \end{subequations} \begin{equation}
E_{\mathbf{k}}=J\sqrt{1+4\cos(\frac{k_{x}l}{2})\left[\cos(\frac{\sqrt{3}}{2}k_{y}l)+\cos(\frac{k_{x}l}{2})\right]}\label{2-8}\end{equation}
 with the phase $\varphi_{\mathbf{k}}$ determined by \begin{equation}
\tan2\varphi_{\mathbf{k}}=-\frac{2\cos\left(k_{x}l/2\right)\sin\left(\sqrt{3}k_{y}l/2\right)}{1+2\cos\left(k_{x}l/2\right)\cos\left(\sqrt{3}k_{y}l/2\right)}.\label{2--9}\end{equation}
 We have to point out that the energy $2E_{\mathbf{k}}$ of \ single
electron excitation actually has six Dirac points on the six vertices
of the first Brillouin Zone in the momentum space. It has been discovered
that in the vicinity of Dirac points, the effective motion of the
electrons accords with the relativistic theory, which is described
by the massless or massive Dirac equation with an effective light
velocity .

In order to study the quantum optical properties of the nanotubes,
it is necessary to introduce a quantized light field \begin{equation}
H_{p}=\sum\limits _{\mathbf{q}}\hbar\Omega_{\mathbf{q}}d_{\mathbf{q}}^{\dagger}d_{\mathbf{q}},\label{2-10}\end{equation}
 where $\Omega_{\mathbf{q}}$ is the frequency of photons with momentum
$\mathbf{q}$. $d_{\mathbf{q}}^{\dagger}$ and $d_{\mathbf{q}}$ creates
and annihilates a photon with momentum $\mathbf{q},$ respectively$.$
We choose $\hbar=1$ and only one polarization direction for each
mode of light denoted by $\mathbf{q}$ in the following discussions.

The interaction between the carbon nanotube and the light field is
obtained according to the minimal coupling principle of
electromagnetic field. By replacing the mechanical momentum of the
electrons with canonical ones and neglecting the multi-photon
interactions, the interaction Hamiltonian is obtained as
\begin{equation} H_{I}=-\frac{e}{mc}\sum\limits
_{\mathbf{k},\mathbf{q}}\mathbf{k}\cdot\mathbf{A}_{\mathbf{q}}\left(a_{\mathbf{k}}^{\dagger}+b_{\mathbf{k}}^{\dagger}\right)\left(a_{\mathbf{k-q}}+b_{\mathbf{k-q}}\right).\label{2-11}\end{equation}
 Here, the vector potential of the quantized light field is \begin{equation}
\mathbf{A}_{\mathbf{q}}=-i\sqrt{\frac{1}{2\epsilon_{0}V\Omega_{\mathbf{q}}}}\mathbf{e}_{\mathbf{q}}\left(d_{\mathbf{q}}-d_{-\mathbf{q}}^{\dagger}\right),\label{2-12}\end{equation}
 where $\mathbf{e}_{\mathbf{q}}$ is the unit polarization vector
of mode $\mathbf{q}.$ $\epsilon_{0}$ is the vacuum electric permittivity
and $V$ is the volume effectively occupied by the light field.

So far, we have obtained the quantized mode of the SWCNT interacting
with a light field, whose Hamiltonian is $H=H_{e}+H_{p}+H_{I}$, with
\begin{subequations} \begin{eqnarray}
H_{e} & = & \sum\limits _{\mathbf{k}}E_{\mathbf{k}}\left(\alpha_{\mathbf{k}}^{\dagger}\alpha_{\mathbf{k}}-\beta_{\mathbf{k}}^{\dagger}\beta_{\mathbf{k}}\right),\label{2-13-1}\\
H_{p} & = & \sum\limits _{\mathbf{q}}\hbar\Omega_{\mathbf{q}}d_{\mathbf{q}}^{\dagger}d_{\mathbf{q}},\label{2-13-2}\\
H_{I} & = & \sum_{\mathbf{k},\mathbf{q}}D_{\mathbf{k,q}}\left(d_{\mathbf{q}}\alpha_{\mathbf{k}}^{\dagger}\beta_{\mathbf{k}-\mathbf{q}}+h.c.\right),\label{2-13-3}\end{eqnarray}
 where we have made the rotating wave approximation to eliminate the
fast varying terms, such as $d_{-\mathbf{q}}^{\dagger}\alpha_{\mathbf{k}}^{\dagger}\beta_{\mathbf{k-q}}$,
$d_{\mathbf{q}}\beta_{\mathbf{k}}^{\dagger}\alpha_{\mathbf{k-q}}$,
$d_{-\mathbf{q}}^{\dagger}\alpha_{\mathbf{k}}^{\dagger}\alpha_{\mathbf{k-q}}$,
$d_{\mathbf{q}}\alpha_{\mathbf{k}}^{\dagger}\alpha_{\mathbf{k-q}}$,
$d_{-\mathbf{q}}^{\dagger}\beta_{\mathbf{k}}^{\dagger}\beta_{\mathbf{k-q}}$,
and $d_{\mathbf{q}}\beta_{\mathbf{k}}^{\dagger}\beta_{\mathbf{k-q}}$,
and the coefficient $D_{\mathbf{k,q}}$ for electron-photon interaction
is \end{subequations} \begin{eqnarray}
D_{\mathbf{k,q}} & = & -\frac{e}{\sqrt{2}mc}\mathbf{k}\cdot\mathbf{e}_{\mathbf{q}}\sqrt{\frac{\hbar}{2\epsilon_{0}V\Omega_{\mathbf{q}}}}(\cos\varphi_{\mathbf{k}}\sin\varphi_{\mathbf{k-q}}\notag\\
 &  & +\cos\varphi_{\mathbf{k-q}}\sin\varphi_{\mathbf{k}}).\label{2-14}\end{eqnarray}

We note that when interaction between the light field and the SWCNT
is significant, the momentum of photons in the light field is
approximately $\left\vert \mathbf{q}\right\vert \sim10^{7}m^{-1}$,
which is much smaller than the momentum of the electron near the
boundary of the first Brillouin Zone of graphene $\left\vert
\mathbf{k}\right\vert \sim10^{10}m^{-1}.$ Thus we neglect the
momentum $\mathbf{q}$ of photons so that
$\cos\varphi_{\mathbf{k}}\sin\varphi_{\mathbf{k-q}}\approx\cos\varphi_{\mathbf{k-q}}\sin\varphi_{\mathbf{k}}\approx\sin[2\varphi_{\mathbf{k}}]/2$,
and the coefficient $D_{\mathbf{k,q}}$ is approximately
\begin{equation}
D_{\mathbf{k,q}}=-\frac{e}{\sqrt{2}mc}\mathbf{k}\cdot\mathbf{e}_{\mathbf{q}}\sqrt{\frac{\hbar}{2\epsilon_{0}V\Omega_{\mathbf{q}}}}\sin2\varphi_{\mathbf{k}}.\label{2-15}\end{equation}
Specially, $D_{\mathbf{k,q}}$ is taken average over all polarization
directions of the light field to obtain the final $D_{\mathbf{k,q}}$
we use in calculations.


%
\begin{figure}[ptb]
\begin{centering}
\includegraphics[bb=39 81 400 569,clip,width=3.5in]{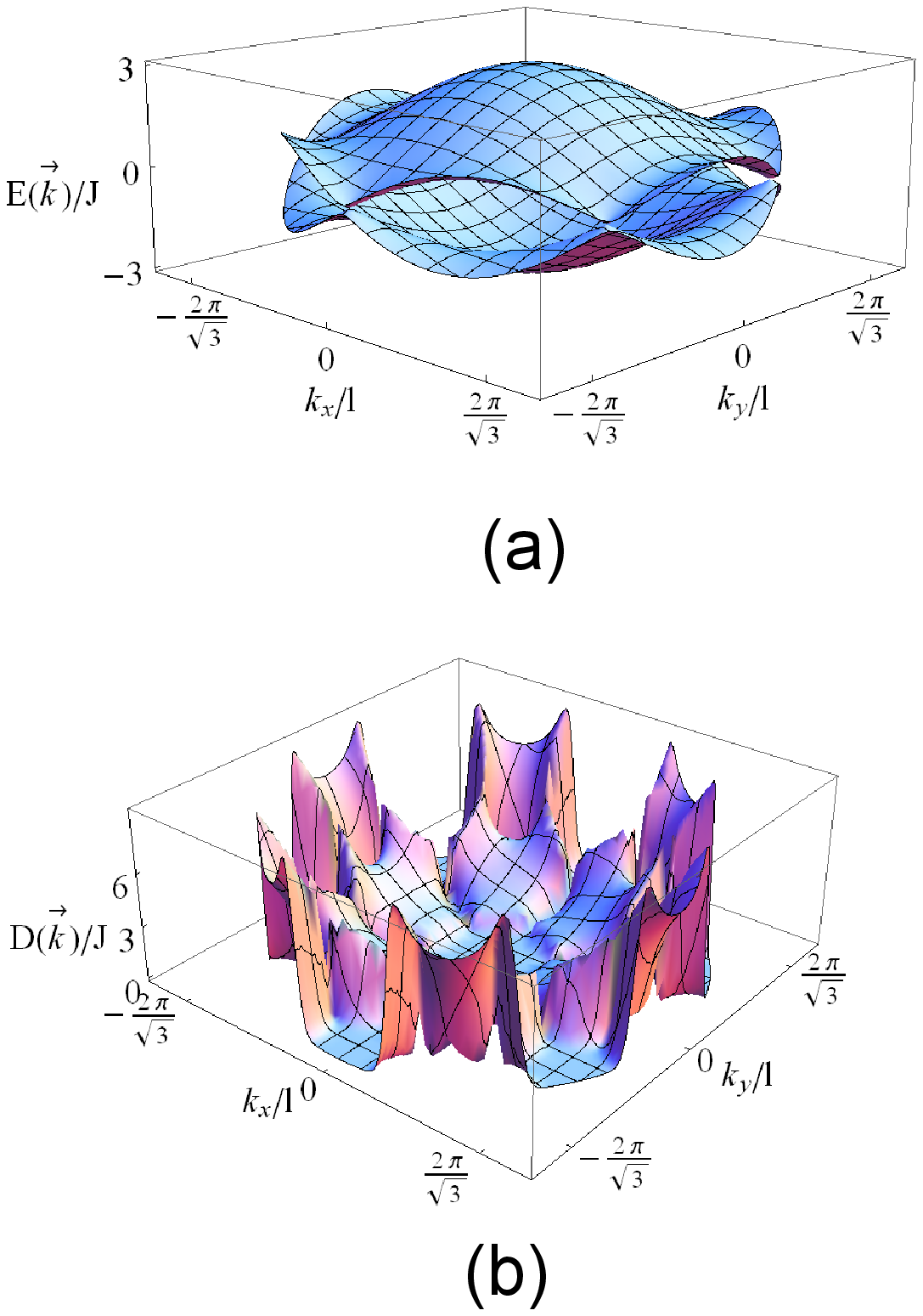}
\par\end{centering}

\caption{(a)The energy spectrum $E(k)$ of graphene versus $k$.
(b)The interaction intensity $D(k)$ between electrons in graphene
and single-mode light, in which we take the average over all the
possible directions for e(q).}

\label{fig:energy}
\end{figure}


The single quasi-particle energy $E_{\mathbf{k}}$ and the interaction
coefficient $D_{\mathbf{k,q}}$ are plotted versus the momentum $\mathbf{k}$
of the electrons in Fig. \ref{fig:energy}. The six Dirac points are
clear to be found at the degeneracy points of upper and lower bands
in Fig. \ref{fig:energy}(a). For the photon momentum $\left\vert \mathbf{q}\right\vert \mathbf{\ll\left\vert \mathbf{k}\right\vert }$
chosen in Fig. \ref{fig:energy}(b), the absolute value of the interaction
coefficient $D_{\mathbf{k,q}}$ becomes large when $\mathbf{k}$ is
near the boundary of the first Brillouin Zone and decreases rapidly
as $\mathbf{k}$ deviate from that boundary.

\section{\label{sec:Interband}Interband Rabi Oscillation Induced by Strong
Light Field}

The general photon-electron interaction contains multi-mode light
field, which case is too complex to be analytically treated in
revealing the essential properties. Thus, we simplify the
Hamiltonian by making the reasonable assumption that only one
particular quantum mode of the light field would dominate the
dynamics. This could be experimentally realized by adding a
high-finesse microcavity to the system to pick out a single mode of
quantized light under consideration. In this sense, the model
Hamiltonian is reduced to $H=H_{0}+H_{1},$ where

\begin{subequations} \begin{eqnarray}
H_{0} & = & \sum\limits _{\mathbf{k}}E_{\mathbf{k}}\left(\alpha_{\mathbf{k}}^{\dagger}\alpha_{\mathbf{k}}-\beta_{\mathbf{k}}^{\dagger}\beta_{\mathbf{k}}\right)+\Omega d^{\dagger}d,\label{3-1-2}\\
H_{1} & = & \sum_{\mathbf{k}}D_{\mathbf{k}}\left(d\alpha_{\mathbf{k}}^{\dagger}\beta_{\mathbf{k-q}}+h.c.\right),\label{3-1-3}\end{eqnarray}
 indicates that a single-mode light field would induce the coherent
transitions of electrons between the upper band and the lower band.
The output of the electronic flow would display an obvious resonance,
namely, Rabi oscillation, which is experimentally observable.

In the strong light limit, the light field can be treated as a classical
one, where the creation and annihilation operators $d^{\dagger}$
and $d$ are replaced by C-numbers, namely \end{subequations} \begin{equation}
d\rightarrow\sqrt{N}e^{-i\Omega t},d^{\dagger}\rightarrow\left(d\right)^{\ast}.\label{3-2}\end{equation}
 with $N$ the total number of photons. This approximation is valid since
in a strong light field only the intensity of the light plays an
important role. We can generally prove this classical approximation
in App.~\ref{App:strong}.

Then we obtain the semi-classical Hamiltonian $H(t)=\sum\limits _{\mathbf{k}}h_{\mathbf{k}}(t),$
in which the single momentum Hamiltonian is \begin{eqnarray}
h_{\mathbf{k}}(t) & = & E_{\mathbf{k}}\left(\alpha_{\mathbf{k}}^{\dagger}\alpha_{\mathbf{k}}-\beta_{\mathbf{k}-\mathbf{q}}^{\dagger}\beta_{\mathbf{k}-\mathbf{q}}\right)+\notag\\
 &  & \sqrt{N}D_{\mathbf{k}}\left(\alpha_{\mathbf{k}}^{\dagger}\beta_{\mathbf{k}-\mathbf{q}}e^{-i\Omega t}+h.c.\right).\label{3-3}\end{eqnarray}
 for electrons with momentum $\mathbf{k}$. Here, we have neglected
the constant $N\Omega$ in the total energy of the light field and
the difference between $E_{\mathbf{k}}$ and $E_{\mathbf{k-q}}$ for
the reason mentioned at the end of Sec.~\ref{sec:setup}. In terms of
the quasi-spin operators \begin{subequations} \begin{eqnarray}
S_{\mathbf{k}}^{z} & = & \frac{1}{2}\left(\alpha_{\mathbf{k}}^{\dagger}\alpha_{\mathbf{k}}-\beta_{\mathbf{k}-\mathbf{q}}^{\dagger}\beta_{\mathbf{k}-\mathbf{q}}\right),\label{3-4-1}\\
S_{\mathbf{k}}^{+} & = & \alpha_{\mathbf{k}}^{\dagger}\beta_{\mathbf{k}-\mathbf{q}},S_{\mathbf{k}}^{-}=\left(S_{\mathbf{k}}^{+}\right)^{\dagger},\label{3-4-2}\end{eqnarray}
which obviously satisfy the commutation relations of the regular spin-$1/2$
operators, the above single momentum Hamiltonian is rewritten as

\end{subequations} \begin{equation}
h_{\mathbf{k}}(t)=E_{\mathbf{k}}S_{\mathbf{k}}^{z}+\sqrt{N}D_{\mathbf{k}}\left(S_{\mathbf{k}}^{+}e^{-i\Omega t}+h.c.\right).\label{3-5}\end{equation}
 It describes a quasi-spin precession in a time-dependent effective
magnetic field\begin{equation}
\mathbf{B=(}\sqrt{N}D_{\mathbf{k}}\cos\Omega t,\sqrt{N}D_{\mathbf{k}}\sin\Omega t,E_{\mathbf{k}}\mathbf{).}\label{3-6}\end{equation}
 Such spin precession is just the Rabi oscillation between bands.

To solve the dynamic equation governed by $h_{\mathbf{k}}(t),$ a
time-dependent unitary transformation \begin{equation}
U(t)=\exp(i\Omega S_{\mathbf{k}}^{z}t),\label{3-7}\end{equation}
 is used to transform the Hamiltonian above into a time-independent
one $h_{\mathbf{k}}^{\prime}=U^{\dagger}h_{\mathbf{k}}(t)U-i\partial_{t}U^{\dagger}U$
or \begin{equation}
h_{\mathbf{k}}^{\prime}=-\Delta_{\mathbf{k}}S_{\mathbf{k}}^{z}+\sqrt{N}D_{\mathbf{k}}S_{\mathbf{k}}^{+}+h.c\label{3-8}\end{equation}
 Here, \begin{equation}
\Delta_{\mathbf{k}}=\Omega-2E_{\mathbf{k}}.\label{3-9}\end{equation}
 is the detuning between the energy of the light field and that of
the quasi-spin.

The Heisenberg equations of the system \begin{subequations} \begin{eqnarray}
i\frac{\partial}{\partial t}S_{\mathbf{k}}^{z} & = & \frac{1}{2}\varepsilon_{\mathbf{k}}\sin\theta_{\mathbf{k}}[S_{\mathbf{k}}^{+}-S_{\mathbf{k}}^{-}],\label{3-10-1}\\
i\frac{\partial}{\partial t}S_{\mathbf{k}}^{\pm} & = & \pm\varepsilon_{\mathbf{k}}[\cos\theta_{\mathbf{k}}S_{\mathbf{k}}^{\pm}+\sin\theta_{\mathbf{k}}S_{\mathbf{k}}^{z}],\label{3-10-2}\end{eqnarray}
 determine the Rabi oscillation of the electrons with momentum $\mathbf{k}$
between the upper and lower bands. Here the mixing angle $\theta_{\mathbf{k}}$
is defined by \end{subequations} \begin{subequations} \begin{eqnarray}
\cos\theta_{\mathbf{k}} & = & \frac{\Delta_{\mathbf{k}}}{\varepsilon_{\mathbf{k}}},\label{3-11-1}\\
\sin\theta_{\mathbf{k}} & = & \frac{2\sqrt{N}D_{\mathbf{k}}}{\varepsilon_{\mathbf{k}}},\label{3-11-2}\\
\varepsilon_{\mathbf{k}}^{2} & = & \Delta_{\mathbf{k}}^{2}+4ND_{\mathbf{k}}^{2}.\label{3-11-3}\end{eqnarray}
 The above first order partial differential equations (\ref{3-10-1})-(\ref{3-10-2})
with initial operators $S_{\mathbf{k}}^{z}(0)$ and $S_{\mathbf{k}}^{\pm}(0)$
is solved through the Laplace transformation \end{subequations} \begin{equation}
\lambda(p)=\int\limits _{0}^{+\infty}\lambda(t)e^{-pt}dt,\label{3-12}\end{equation}
 which gives \begin{subequations} \begin{eqnarray}
pS_{\mathbf{k}}^{z}-S_{\mathbf{k}}^{z}(0) & = & -i\frac{1}{2}\varepsilon_{\mathbf{k}}\sin\theta_{\mathbf{k}}[S_{\mathbf{k}}^{+}-S_{\mathbf{k}}^{-}],\label{3-13-1}\\
pS_{\mathbf{k}}^{\pm}-S_{\mathbf{k}}^{\pm}(0) & = & \mp i\varepsilon_{\mathbf{k}}[\cos\theta_{\mathbf{k}}S_{\mathbf{k}}^{\pm}+\sin\theta_{\mathbf{k}}S_{\mathbf{k}}^{z}].\label{3-13-2}\end{eqnarray}
 In terms of the normalized Laplacian parameter $p^{\prime}=p/\varepsilon_{\mathbf{k}},$
the above equation is solved as \
\end{subequations} \begin{subequations} \begin{eqnarray}
S_{\mathbf{k}}^{z}(p^{\prime}) & = & \frac{S_{\mathbf{k}}^{z}(0)}{\varepsilon_{\mathbf{k}}}\frac{[p^{\prime2}+\cos^{2}\theta_{\mathbf{k}}]}{p^{\prime}[p^{\prime2}+1]}+\frac{S_{\mathbf{k}}^{y}(0)}{\varepsilon_{\mathbf{k}}}\frac{\sin\theta_{\mathbf{k}}}{p^{\prime2}+1}\notag\nonumber \\
 &  & -\frac{S_{\mathbf{k}}^{x}(0)}{\varepsilon_{\mathbf{k}}}\frac{\sin\theta_{\mathbf{k}}\cos\theta_{\mathbf{k}}}{p^{\prime}[p^{\prime2}+1]},\label{3-14-1}\\
S_{\mathbf{k}}^{\pm}(p^{\prime}) & = & \frac{1}{\varepsilon_{\mathbf{k}}}\frac{S_{\mathbf{k}}^{\pm}(0)\mp i\sin\theta_{\mathbf{k}}S_{\mathbf{k}}^{z}(p^{\prime})}{p^{\prime}\pm i\cos\theta_{\mathbf{k}}},\label{3-14-2}\end{eqnarray}
 where \end{subequations} \begin{subequations} \begin{eqnarray}
S_{\mathbf{k}}^{x} & = & \frac{1}{2}(S_{\mathbf{k}}^{+}+S_{\mathbf{k}}^{-}),\label{3-15-1}\\
S_{\mathbf{k}}^{y} & = & \frac{1}{2i}(S_{\mathbf{k}}^{+}-S_{\mathbf{k}}^{-}).\label{3-15-2}\end{eqnarray}
 In the SWCNT, the electrons fill up the lower band when the system
stays at its ground state at zero temperature. As a consequence, we
may simply set $S_{\mathbf{k}}^{x}(0)$ and $S_{\mathbf{k}}^{y}(0)$
as zero for convenience in the following discussions. The inverse
Laplace transformation gives the time evolution of $S_{\mathbf{k}}^{z}(t)$
and $S_{\mathbf{k}}^{\pm}(t)$ respectively \end{subequations} \begin{equation}
S_{\mathbf{k}}^{z}(t)=S_{\mathbf{k}}^{z}(0)\left[\cos^{2}\theta_{\mathbf{k}}+\sin^{2}\theta_{\mathbf{k}}\cos\left(\varepsilon_{\mathbf{k}}t\right)\right].\label{3-16}\end{equation}
 and \begin{eqnarray}
S_{\mathbf{k}}^{\pm}(t) & = & -S_{\mathbf{k}}^{z}(0)\sin\theta_{\mathbf{k}}\cos\theta_{\mathbf{k}}[1-\cos\left(\varepsilon_{\mathbf{k}}t\right)]\notag\nonumber \\
 &  & \mp iS_{\mathbf{k}}^{z}(0)\sin\theta_{\mathbf{k}}\sin\left(\varepsilon_{\mathbf{k}}t\right).\label{3-17}\end{eqnarray}

Finally, the total population inversion \begin{equation}
W(t)=\sum\limits _{\mathbf{k}}\left\langle S_{\mathbf{k}}^{z}(t)+\frac{1}{2}\right\rangle \label{3-18}\end{equation}
 is calculated as the summation over those of\ single momentum, which
reads\begin{equation}
W(t)=\sum\limits _{\mathbf{k}}\left\{ \left\langle S_{\mathbf{k}}^{z}(0)\right\rangle \left[1-2\sin^{2}\theta_{\mathbf{k}}\sin^{2}\left(\frac{\varepsilon_{\mathbf{k}}t}{2}\right)\right]+\frac{1}{2}\right\} .\label{3-19}\end{equation}
 When the temperature is zero, the system stays at its ground state
and thus $\left\langle S_{\mathbf{k}}^{z}(t=0)\right\rangle =-1/2$
is valid for all $\mathbf{k}$. Then the total population inversion
is obtained \begin{equation}
W(t)=\sum\limits _{\mathbf{k}}\frac{1}{2}\sin^{2}\theta_{\mathbf{k}}\{1-\cos\left(\varepsilon_{\mathbf{k}}t\right)\}\label{3-20}\end{equation}

If we consider the continuous momentum in a 2-D graphene and the inhomogeneously-broadened
system in which different quasi-spins have different momentums by
introducing the distribution $g(\varepsilon_{\mathbf{k}})$ centered
on $\varepsilon_{\mathbf{0}}$ as\begin{equation}
g(\varepsilon_{\mathbf{k}})=2\sqrt{\pi}T\exp\left[-T^{2}\left(\varepsilon_{\mathbf{k}}-\varepsilon_{\mathbf{0}}\right)^{2}\right],\label{3-21}\end{equation}
 which satisfies $\frac{1}{2\pi}\int_{\infty}^{-\infty}g(\varepsilon_{\mathbf{k}})d\varepsilon_{\mathbf{k}}=1$.
When $2\sqrt{N}D_{\mathbf{k}}\gg\Delta_{\mathbf{k}}$ results in $\sin\theta_{\mathbf{k}}\simeq1$,
the total population inversion can be calculated as\begin{eqnarray}
W(t) & = & \frac{1}{4\pi}\int_{\infty}^{-\infty}g(\varepsilon_{\mathbf{k}})\{1-\cos\left(\varepsilon_{\mathbf{k}}t\right)\}d\varepsilon_{\mathbf{k}}\mbox{\notag}\\
 & = & \frac{1}{2}\left[1-\cos\left(\varepsilon_{\mathbf{0}}t\right)\right]\exp\left(-\frac{t^{2}}{4T^{2}}\right).\label{3-22}\end{eqnarray}
 It must be pointed out that the time dependence of the total population
inversion includes two aspects when the energy distribution is Gaussian
type. One is the periodic factor as $(1-\cos\left(\varepsilon_{\mathbf{0}}t\right))$
resulting from the central frequency of the Gaussian distribution.
The other is the exponential decay $\exp\left(-\frac{t^{2}}{4T^{2}}\right)$
resulted from the broadening of the Gaussian distribution. The randomness
of the energy spectrum of the quasi-spins actually induces these effects,
which can be considered as a kind of spin echo.


%
\begin{figure}[ptb]

\begin{centering}
\includegraphics[bb=15 18 311 246,clip,width=3.5in]{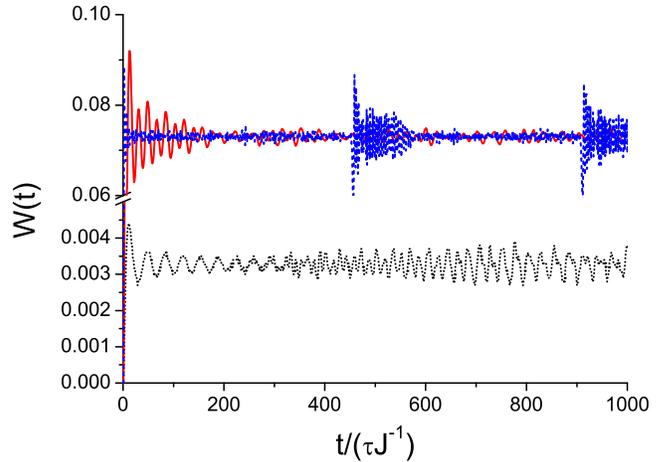}
\par\end{centering}

\caption{Population inversion of electrons in the SWCNT is plotted
versus time. The parameters for the SWCNT are respectively:
(1)chiral vector is $(6,4)$, $\Omega=0.4$, $\tau=5$ for the black
short dotted line; (2)chiral vector is $(8,0)$, $\Omega=2$, $\tau=5$
for the red solid line; (3)chiral vector is $(8,0)$, $\Omega=2$,
$\tau=70$ for the blue short dashed line. Here, $\tau$ is the time
scale.}

\label{fig:popinversion}
\end{figure}


From Fig.~\ref{fig:popinversion} for the population inversion of
the $(2n,0)$ SWCNTs with the incurring light frequency of $\Omega=2J$,
we may see that a considerable proportion of the electrons are excited
to the upper band (more than $1/15$ for $(8,0)$ SWCNT), and exhibits
collapse and revival in a long period of time. The explanation for
it is straightforward: in the $(2n,0)$ SWCNTs, there are large degeneracies
of possible states onto the equi-energy lines $E=J$ of the $2$-D
graphene energy bands. Thus, the $(2n,0)$ SWCNTs are potential experimental
candidates for the demonstration of Rabi oscillation in solids.

\section{\label{sec:interference}First Order Coherence of Scattered and Emitted
Photons}

The strong light field only couples the upper and lower bands of electrons
through its intensity, which essentially cancels the quantum optical
features of the SWCNT characterized by the higher order quantum coherence.
To save curiosity of quantized light field interacting with SWCNT,
we return to the Hamiltonian Eq.(\ref{3-1-2})-(\ref{3-1-3})

The first order correlation function of the light field
\begin{equation} G^{(1)}(\tau)=\left\langle
d^{+}(t)d(t+\tau)\right\rangle \label{4-1}\end{equation} to
characterize the interference of the electrons in SWCNT is
independent of $t$ after long time evolution $t\rightarrow+\infty$,
which corresponds to the steady solution for the light-SWCNT
coupling system. In the interaction picture with respect
to\begin{equation} H_{0}=\Omega d^{\dagger}d+\sum\limits
_{\mathbf{k}}\Omega S_{\mathbf{k}}^{z},\label{4-2}\end{equation}
 the Langevin equations read as \begin{subequations} \begin{eqnarray}
\frac{\partial}{\partial t}d & = & -i\sum\limits _{\mathbf{k}}D_{\mathbf{k}}S_{\mathbf{k}}^{-},\label{4-3-1}\\
\frac{\partial}{\partial t}S_{\mathbf{k}}^{-} & = & \left(i\Delta_{\mathbf{k}}-\gamma_{\mathbf{k}}\right)S_{\mathbf{k}}^{-}+2iD_{\mathbf{k}}S_{\mathbf{k}}^{z}d,\label{4-3-2}\\
\frac{\partial}{\partial t}S_{\mathbf{k}}^{z} & = & -2\gamma_{\mathbf{k}}(S_{\mathbf{k}}^{z}+\frac{1}{2})+iD_{\mathbf{k}}(d^{\dagger}S_{\mathbf{k}}^{-}-S_{\mathbf{k}}^{+}d).\label{4-3-3}\end{eqnarray}
 Here, we phenomenologically add decay terms of the SWCNT part to
the Langevin equations, while neglect the decay of light field since
an ideal probe is considered. We also assume that the SWCNT system
reaches its equilibrium state with the light field before there is
considerable change in the light field. Actually, this assumption is
very crucially used in Haken's theory of laser \cite{Laser}. By
setting the time derivatives of the $S$ operators as zero, the
steady solution of the total system can be obtained with steady
quasi-spin operators \end{subequations}
\begin{subequations}
\begin{eqnarray}
S_{\mathbf{k}}^{z} & = & -\frac{\gamma_{\mathbf{k}}^{2}+\Delta_{\mathbf{k}}^{2}}{2(\Delta_{\mathbf{k}}^{2}+2d^{\dagger}dD_{\mathbf{k}}^{2}+\gamma_{\mathbf{k}}^{2})},\label{4-4-1}\\
S_{\mathbf{k}}^{-} & = & -\frac{iD_{\mathbf{k}}(\gamma_{\mathbf{k}}-i\Delta_{\mathbf{k}})d}{\Delta_{\mathbf{k}}^{2}+2d^{\dagger}dD_{\mathbf{k}}^{2}+\gamma_{\mathbf{k}}^{2}}.\label{4-4-2}\end{eqnarray}
 Therefore, if the number of photons does not fluctuate intensively
long time after the light is turned on, we could simply set the
particle number operator $d^{\dagger}d=N$ as a constant.

In order to study the first order coherence of the light field, we
use the mean field approach for the Langevin equations of the above
system by setting \end{subequations} \begin{equation}
S_{\mathbf{k}}^{z}d\approx\left.\left\langle
S_{\mathbf{k}}^{z}(t)\right\rangle \right\vert
_{t\rightarrow\infty}d(\tau)\equiv
S_{\mathbf{k}}^{z}(\infty)d(\tau)\label{4-5}\end{equation}
 for long time evolution. Here we can analytically calculate the first
order correlation function through the partial differential equations
(\ref{4-3-1}-\ref{4-3-3}). After applying Laplace transformation
to Eq. (\ref{4-3-1}-\ref{4-3-3}), we have \begin{subequations}
\begin{eqnarray}
pd-d(0) & = & -i\sum\limits _{\mathbf{k}}D_{\mathbf{k}}S_{\mathbf{k}}^{-},\label{4-6-1}\\
\left(p-i\Delta_{\mathbf{k}}^{\prime}\right)S_{\mathbf{k}}^{-} & = & 2iD_{\mathbf{k}}\left.\left\langle S_{\mathbf{k}}^{z}(t)\right\rangle \right\vert _{t\rightarrow\infty}d+S_{\mathbf{k}}^{-}(0),\label{4-6-2}\end{eqnarray}
 for the effective detuning $\Delta_{\mathbf{k}}^{\prime}=\Delta_{\mathbf{k}}+i\gamma_{\mathbf{k}}.$
This gives the solution of $d(p)$ as \end{subequations} \begin{equation}
d(p)=\frac{d(0)+\Lambda^{-}(p)}{p-\Lambda^{z}(p)}.\label{4-7}\end{equation}
 Here, \begin{equation}
\Lambda^{-}(p)=-i\sum\limits _{\mathbf{k}}\frac{D_{\mathbf{k}}S_{\mathbf{k}}^{-}(0)}{p-i\Delta_{\mathbf{k}}^{\prime}}\label{4-8}\end{equation}
 represents the contribution from $S_{\mathbf{k}}^{-}(0)$, while
contribution from the long time evolution of $\left\langle S_{\mathbf{k}}^{z}\right\rangle $
is given by \begin{equation}
\Lambda^{z}(p)=\sum\limits _{\mathbf{k}}\frac{2D_{\mathbf{k}}^{2}\left.\left\langle S_{\mathbf{k}}^{z}(t)\right\rangle \right\vert _{t\rightarrow\infty}}{p-i\Delta_{\mathbf{k}}^{\prime}}.\label{4-9}\end{equation}

Since the electron-photon interaction serves as a perturbation term
in the Hamiltonian, the singularities of the $d(p)$ is mainly
determined by the denominator $p-\Lambda^{z}(p)$. Under the
Wigner-Weisskopf approximation, the $0$-th order zero point of the
denominator is $p=0,$ and to the $1$-st order it is
\begin{equation}
p=-i\Omega^{\prime}-\Gamma^{\prime},\label{4-10}\end{equation}
 where the renormalized frequency and the effective decay are \begin{subequations}
\begin{eqnarray}
\Omega^{\prime} & = & -\mathrm{Im}\Lambda^{z}(0),\label{4-11-1}\\
\Gamma^{\prime} & = & -\mathrm{Re}\Lambda^{z}(0).\label{4-11-2}\end{eqnarray}
 Applying the inverse Laplace transformation, we obtain an expression
for $d(\tau)$ as \end{subequations} \begin{equation}
d(\tau)=\exp\left(-i\Omega^{\prime}\tau-\Gamma^{\prime}\tau\right)\left[d(0)+F(\tau)\right],\label{4-12}\end{equation}
 where the contribution from $S_{\mathbf{k}}^{-}(0)$ is
\begin{equation}
F(\tau)=-i\sum\limits
_{\mathbf{k}}\frac{D_{\mathbf{k}}S_{\mathbf{k}}^{-}(0)}{\mu_{\mathbf{k}}
+i\nu_{\mathbf{k}}}\left[1-e^{-i\mu_{\mathbf{k}}\tau-\nu_{\mathbf{k}}\tau}\right],\label{4-13}
\end{equation}
 with $\mu_{\mathbf{k}}=\left(-\Delta_{\mathbf{k}}-\Omega^{\prime}\right)$
and $\nu_{\mathbf{k}}=\left(\gamma_{\mathbf{k}}-\Gamma^{\prime}\right).$
If we compare a quasi-spin system to a heat bath, the term $F(\tau)$
represents its induced quantum fluctuation. The couplings of the light
field to SWCNT is characterized by \begin{subequations} \begin{eqnarray}
\Omega^{\prime} & = & \sum\limits _{\mathbf{k}}\frac{D_{\mathbf{k}}^{2}}{\Delta_{\mathbf{k}}^{2}+2ND_{\mathbf{k}}^{2}+\gamma_{\mathbf{k}}^{2}}\Delta_{\mathbf{k}},\label{4-14-1}\\
\Gamma^{\prime} & = & \sum\limits _{\mathbf{k}}\frac{D_{\mathbf{k}}^{2}}{\Delta_{\mathbf{k}}^{2}+2ND_{\mathbf{k}}^{2}+\gamma_{\mathbf{k}}^{2}}\gamma_{\mathbf{k}},\label{4-14-2}\end{eqnarray}
 where the explicit steady solution for $S_{\mathbf{k}}^{z}$ that
 has been used.\end{subequations}

The contribution $\left\langle d^{+}(0)F(\tau)\right\rangle $ from
$S_{\mathbf{k}}^{-}(0)$ in the first order correlation \begin{equation}
G^{(1)}(\tau)=\exp\left(-i\Omega^{\prime}\tau-\Gamma^{\prime}\tau\right)\left(G^{(1)}(0)+\left\langle d^{+}(0)F(\tau)\right\rangle \right)\label{4-15}\end{equation}
 vanishes since the average on the photon number states reduces to zero due to
the photon number conservation. Then the normalized first order coherence
function $g^{(1)}(\tau)\equiv G^{(1)}(\tau)/G^{(1)}(0)$ is explicitly
written as \begin{equation}
g^{(1)}(\tau)=\exp\left(-i\Omega^{\prime}\tau-\Gamma^{\prime}\tau\right),\label{4-16}\end{equation}
 which is used to measure the interference of the scattered and emitted
photons. It is clear that long time first order correlation of the
light field vanishes exponentially with $\tau$. In most cases,
considering that the decay rates $\gamma_{\mathbf{k}}$ are much
smaller than the detuning $\Delta_{\mathbf{k}}$,
$\Gamma^{\prime}\ll\Omega^{\prime}$ is obviously satisfied. Thus,
neglecting the decay effect in the first order coherence, the
existence of SWCNT contributes to the first order coherence a shift
$\Omega^{\prime}$ in the frequency of light.

\section{\label{sec:Second}Second Order Correlation of The Scattered and
Emitted Light}

The first order coherence function only demonstrates the interference
of the scattered and emitted photon. To distinguish the fully quantum
optical properties of the SWCNT, e.g. , the bunching and anti-bunching
of the photons, the second order coherence function \begin{equation}
G^{(2)}(\tau)=\left\langle d^{\dagger}(t)d^{\dagger}(t+\tau)d(t+\tau)d(t)\right\rangle \label{5-1}\end{equation}
 is needed. According to Eq. (\ref{4-10}), we calculate\begin{eqnarray}
G^{(2)}(\tau) & = & \exp\left(-2\Gamma^{\prime}\tau\right)G^{(2)}(0)\notag\\
 &  & +\exp\left(-2\Gamma^{\prime}\tau\right)\left\langle d^{\dagger}(0)F^{\dagger}(\tau)F(\tau)d(0)\right\rangle ,\label{5-2}\end{eqnarray}
 where we have neglected terms $\left\langle d^{\dagger}(0)d^{\dagger}(0)F(\tau)d(0)\right\rangle $
and $\left\langle d^{\dagger}(0)F^{\dagger}(\tau)d(0)d(0)\right\rangle $
because of the photon number conservation for the light field. Neglecting
correlation between different quasi-spins, only terms with the same
momentum can survive in $F^{\dagger}(\tau)$ and $F(\tau).$ Therefore,
the non-vanishing second term is calculated as \begin{eqnarray}
 &  & \left\langle d^{\dagger}(0)F^{\dagger}(\tau)F(\tau)d(0)\right\rangle \notag\\
 & \approx & \sum\limits _{\mathbf{k}}f_{\mathbf{k}}(\tau)\left\langle d^{\dagger}(0)S_{\mathbf{k}}^{+}(0)S_{\mathbf{k}}^{-}(0)d(0)\right\rangle \notag\\
 & \approx & \sum\limits _{\mathbf{k}}f_{\mathbf{k}}(\tau)\left\langle d^{\dagger}(0)d(0)\right\rangle \left(\left.\left\langle S_{\mathbf{k}}^{z}(t)\right\rangle \right\vert _{t\rightarrow\infty}+\frac{1}{2}\right),\label{5-3}\end{eqnarray}
 where the time dependent coefficients are \begin{equation}
f_{\mathbf{k}}(\tau)=\frac{2D_{\mathbf{k}}^{2}}{\mu_{\mathbf{k}}^{2}+\nu_{\mathbf{k}}^{2}}\left[\cosh\left(\nu_{\mathbf{k}}\tau\right)-\cos\left(\mu_{\mathbf{k}}\tau\right)\right]e^{-\nu_{\mathbf{k}}\tau}.\label{5-4}\end{equation}
 Accordingly, the normalized second order coherence function $g^{(2)}(\tau)\equiv G^{(2)}(\tau)/\left\vert G^{(1)}(0)\right\vert ^{2}$
is written as\begin{eqnarray}
g^{(2)}(\tau) & = & \exp\left(-2\Gamma^{\prime}\tau\right)\left[\frac{G^{(2)}(0)}{G^{(1)}(0)^{2}}\right.+\notag\\
 &  & \left.\sum\limits _{\mathbf{k}}\frac{f_{\mathbf{k}}(\tau)}{G^{(1)}(0)}\left(\left.\left\langle S_{\mathbf{k}}^{z}(t)\right\rangle \right\vert _{t\rightarrow\infty}+\frac{1}{2}\right)\right]\label{5-5}\end{eqnarray}

\begin{figure}[ptb]

\begin{centering}
\includegraphics[bb=15 9 293 236,clip,width=3.5in]{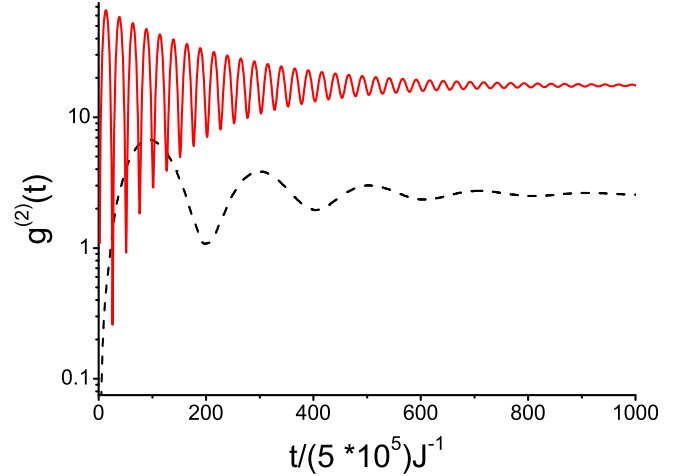}
\par\end{centering}

\caption{The second order correlation of the light
$g^{(2)}(t)-g^{(2)}(0)$ is plotted, in which the antibunching
feature is obviously displayed. The two chiral vectors $(6,4)$ and
$(8,0)$ are chosen to represent significant anti-bunching effect due
to different reasons. Here, we have chosen the frequency of the
light field $\omega=2J$. }

\label{fig:second}
\end{figure}

Here, the second item in $g^{(2)}(\tau)$ is non-negative for any
$\tau$, and returns to zero when $\tau\rightarrow0$, thus \ the
explicit effect of the anti-bunching of the light coupled with the
SWCNT is illustrated in Fig. $~\ref{fig:second}$.


%


Due to the divergence near the resonance area for $g^{(2)}(t)$, the
anti-bunching feature is significant where the light and the energy
gap between upper and lower bands reach resonance while the interaction
intensity $D_{\mathbf{k}}$ is comparatively high. In this case, the
SWCNT is equivalent to one or several 2-level atoms that interact
strongly with the incurring light, just as the case in the $(6,4)$
SWCNT when the incurring light frequency is $2J$. Similar to Sec.~\ref{sec:Interband}
concerning Rabi oscillation, here we still have a distinct effect
for the $(2n,0)$ SWCNTs, when the incurring light frequency is really
close to $2J$. In the case for $(8,0)$ SWCNT, the strong anti-bunching
feature is instead caused by the large degeneracy on the $E=J$ line
in the first Brillouin zone. Unlike the case for the $(6,4)$ SWCNT,
in which merely several electron states are involved, the significant
anti-bunching here is caused by the excitation in the $E=J$ band
of the SWCNT, where thousands of possible states participate in at
the same time.

\section{\label{sec:Possible}Possible Lasing Mechanism of Carbon Nanotube}

The above investigations imply that the light emitted from or
scattered by the SWCNT is strongly correlated in time domain, thus
explicitly displays quantum effects. It is straight forward to
imagine that if electrons in the SWCNT experience a population
inversion, the emitted light would be amplified. This observation
may enable a possible lasing mechanism. In this section, we will
explore this mechanism for the SWCNT by using Haken's laser theory
~\cite{Laser}.

The Heisenberg equations(\ref{3-10-1},\ref{3-10-2}) without
dissipation usually have no steady solution. Thus we
phenomenologically introduce decays on both the light field and the
quasi-spin operators to make the physical observables reach the
stable results. In order to obtain the steady solution, we neglect
the fluctuations because the time average of them vanishes. This
simplification results in the laser-like equations
\begin{subequations}
\begin{align}
\frac{\partial}{\partial t}\widetilde{d}^{\dagger} & = & -\kappa\widetilde{d}^{\dagger}
+i\sum\limits _{\mathbf{k}}D_{\mathbf{k}}\widetilde{S}_{\mathbf{k}}^{+}e^{-i\Delta_{\mathbf{k}}t},\label{6-1-1}\\
\frac{\partial}{\partial t}\widetilde{S}_{\mathbf{k}}^{+} & = & -\gamma_{\mathbf{k}}\widetilde{S}_{\mathbf{k}}^{+}
-2iD_{\mathbf{k}}\widetilde{d}^{\dagger}S_{\mathbf{k}}^{z}e^{i\Delta_{\mathbf{k}}t},\label{6-1-2}\\
\frac{\partial}{\partial t}S_{\mathbf{k}}^{z} & = & -2\gamma_{\mathbf{k}}(S_{\mathbf{k}}^{z}+\frac{1}{2})
-iD_{\mathbf{k}}(\widetilde{S}_{\mathbf{k}}^{+}\widetilde{d}e^{-i\Delta_{\mathbf{k}}t}-h.c.),\label{6-1-3}
\end{align}
\end{subequations}
where we have removed the higher frequency factors by defining
$\widetilde{d}^{\dagger}=d^{\dagger}\exp(-i\Omega t)$ and
$\widetilde{S}_{\mathbf{k}}^{+}=S_{\mathbf{k}}^{+}\exp[-i2E(\overrightarrow{k})t].$
This approach changes the observation from a laboratory frame of
reference into some rotating one. Equation (\ref{6-1-2}) can be
formally integrated as
\begin{equation}
\widetilde{S}_{\mathbf{k}}^{+}(t)=\widetilde{S}_{\mathbf{k}}^{+}(0)e^{-\gamma_{\mathbf{k}}t}
-2iD_{\mathbf{k}}\int\limits_{0}^{t}e^{-\gamma_{\mathbf{k}}(t-\tau)}
\widetilde{d}^{\dagger}S_{\mathbf{k}}^{z}e^{i\Delta_{\mathbf{k}}\tau}d\tau.\label{6-2}
\end{equation}

According to Haken's laser theory, if
$\widetilde{d}^{\dagger}S_{\mathbf{k}}^{z}$ varies with time much
slower than $\widetilde{S}_{\mathbf{k}}^{+}(t)$, it could be
regarded as a time-independent one and then the above integral
becomes\begin{equation}
\widetilde{S}_{\mathbf{k}}^{+}(t)=\widetilde{S}_{\mathbf{k}}^{+}(0)e^{-\gamma_{\mathbf{k}}t}
-2iD_{\mathbf{k}}\widetilde{d}^{\dagger}S_{\mathbf{k}}^{z}\frac{\left(e^{i\Delta_{\mathbf{k}}t}
-e^{-\gamma_{\mathbf{k}}t}\right)}{\gamma_{\mathbf{k}}+i\Delta_{\mathbf{k}}},\label{6-3}
\end{equation}
After a long time, the first term in the above solution
Eq.(\ref{6-2}), which is totally determined by the initial
polarization $\widetilde{S}_{\mathbf{k}}^{+}(0)$, will vanish. Thus,
when $\gamma_{k}t\gg1$, only the initial state-independent part

\begin{equation}
\widetilde{S}_{\mathbf{k}}^{+}(t)\approx-2iD_{\mathbf{k}}\widetilde{d}^{\dagger}S_{\mathbf{k}}^{z}\frac{e^{i\Delta_{\mathbf{k}}t}}{\gamma_{\mathbf{k}}+i\Delta_{\mathbf{k}}},\label{6-4}\end{equation}
 remains. In this case the motion equation
of the $z-$direction spin operators becomes\begin{equation}
\frac{\partial}{\partial t}S_{\mathbf{k}}^{z}\approx-2\gamma_{\mathbf{k}}(S_{\mathbf{k}}^{z}+\frac{1}{2})-\theta_{\mathbf{k}}\widetilde{d}^{\dagger}\widetilde{d}S_{\mathbf{k}}^{z},\label{6-5}\end{equation}
 where $\theta_{\mathbf{k}}=4\gamma_{\mathbf{k}}D_{\mathbf{k}}^{2}/(\gamma_{\mathbf{k}}^{2}+\Delta_{\mathbf{k}}^{2})$.
Then we obtain the effective motion equation of the light
field
\begin{equation} \frac{\partial}{\partial
t}\widetilde{d}^{\dagger}=-\widetilde{d}^{\dagger}\left(\kappa-\sum\limits
_{\mathbf{k}}\frac{2D_{\mathbf{k}}^{2}e^{i\Omega
t}}{\gamma_{\mathbf{k}}+i\Delta_{\mathbf{k}}}S_{\mathbf{k}}^{z}\right).\label{6-6}
\end{equation}
In the following discussions we will demonstrate a lasing-like
phenomenon by considering the solution of Eq.(\ref{6-6})

Usually, a lasing process requires population inversion. To realize
such population inversion in our setup, a pump of electrons is
needed to inject electrons with specific state into the carbon
nanotube. Phenomenologically, we add a pump term $c_{\mathbf{k}}>0$
to each term $S_{\mathbf{k}}^{z}$, then
\begin{equation}
\frac{\partial}{\partial
t}S_{\mathbf{k}}^{z}=c_{\mathbf{k}}-2\gamma_{\mathbf{k}}(S_{\mathbf{k}}^{z}
+\frac{1}{2})-\theta(\mathbf{k})\widetilde{d}^{\dagger}\widetilde{d}S_{\mathbf{k}}^{z},\label{6-7}
\end{equation}
The population inversion is obtained from Eq. (\ref{6-7}) as
\begin{eqnarray}
S_{\mathbf{k}}^{z} & = & S_{\mathbf{k}}^{z}(0)\exp\left(-\int\limits _{0}^{t}
\left[\theta_{\mathbf{k}}\widetilde{d}^{\dagger}\widetilde{d}+2\gamma_{\mathbf{k}}\right]d\tau^{\prime}\right)+\notag\\
&  & (c_{\mathbf{k}}-\gamma_{\mathbf{k}})\int\limits _{0}^{t}\exp
\left(\int\limits _{0}^{\tau}\left[\theta_{\mathbf{k}}\widetilde{d}^{\dagger}\widetilde{d}
+2\gamma_{\mathbf{k}}\right]d\tau^{\prime}\right)d\tau\times\notag\\
&  & \exp\left(-\int\limits
_{0}^{t}\left[\theta_{\mathbf{k}}\widetilde{d}^{\dagger}\widetilde{d}
+2\gamma_{\mathbf{k}}\right]d\tau^{\prime}\right).\label{6-8}\end{eqnarray}
 After a long time evolution $\left(\gamma_{\mathbf{k}}t\gg1\right)$,
this solution becomes \begin{equation}
S_{\mathbf{k}}^{z}=(c_{\mathbf{k}}-\gamma_{\mathbf{k}})\int\limits _{0}^{t}\exp\left(-\int\limits _{\tau}^{t}\theta_{\mathbf{k}}\widetilde{d}^{\dagger}\widetilde{d}d\tau^{\prime}\right)e^{-2\gamma_{\mathbf{k}}(t-\tau)}d\tau.\label{6-9}\end{equation}

It follows from Eq.(\ref{6-9}) that the main contribution of the
integral comes from the accumulation of the weighted photon numbers
in the time $\tau\sim t$. In this sense we can assume that
\begin{equation*}
\int\limits
_{\tau}^{t}\theta_{\mathbf{k}}\widetilde{d^{\dagger}}\widetilde{d}d\tau^{\prime}
=\theta_{\mathbf{k}}\widetilde{d^{\dagger}}\widetilde{d}(t-\tau)
\end{equation*}
Then the population inversion is integrated as
\begin{equation}
S_{\mathbf{k}}^{z}\approx\frac{(c_{\mathbf{k}}-\gamma_{\mathbf{k}})}
{\theta_{\mathbf{k}}\widetilde{d}^{\dagger}\widetilde{d}+2\gamma_{\mathbf{k}}}.\label{6-10}
\end{equation}
Eventually, the motion equation of the light field is obtained as
\begin{equation}
\frac{\partial}{\partial
t}\widetilde{d}^{\dagger}\approx(\kappa^{\prime} -i\delta
\omega)\widetilde{d}^{\dagger}-\eta\widetilde{d}^{\dagger}\widetilde{d}^{\dagger}
\widetilde{d},\label{6-11}
\end{equation}
where
\begin{subequations}
\begin{equation}
\delta \omega=\sum\limits
_{\mathbf{k}}D_{\mathbf{k}}^{2}\frac{(c_{\mathbf{k}}-\gamma_{\mathbf{k}})}
{\gamma_{\mathbf{k}}}\frac{\Delta_{\mathbf{k}}}{\gamma_{\mathbf{k}}^{2}
+\Delta_{\mathbf{k}}^{2}},\label{6-12-1}
\end{equation}
 appears as the Lamb shift of photons, and
 \begin{equation}
\kappa^{\prime}=-\kappa+\sum\limits _{\mathbf{k}}D_{\mathbf{k}}^{2}\frac{(c_{\mathbf{k}}
-\gamma_{\mathbf{k}})}{\gamma_{\mathbf{k}}^{2}+\Delta_{\mathbf{k}}^{2}},\label{6-12-2}
\end{equation}
represents a dissipation or amplification of the optical mode
together with
\begin{equation}
 \eta=\sum\limits
_{\mathbf{k}}2D_{\mathbf{k}}^{4}\frac{c_{\mathbf{k}}-\gamma_{\mathbf{k}}}
{(\gamma_{\mathbf{k}}^{2}+\Delta_{\mathbf{k}}^{2})^{2}}.\label{6-12-3}
\end{equation}
 describing the extent of nonlinearity of the light field induced by
the SWCNT. Here, we have expanded the second item on the right hand
side of Eq.(\ref{6-11}) up to the first order of $2D_{\mathbf{k}}^{2}\widetilde{d}^{\dagger}\widetilde{d}$.

Obviously, Eq.(\ref{6-11}) is typical to describe the lasing process
in an amplification medium. When electrons are injected into the
SWCNT to realize a population inversion, \end{subequations}
\begin{equation} \kappa^{\prime}=-\kappa+\sum\limits
_{\mathbf{k}}D_{\mathbf{k}}^{2}\frac{(c_{\mathbf{k}}-\gamma_{\mathbf{k}})}{\gamma_{\mathbf{k}}^{2}+\Delta_{\mathbf{k}}^{2}}>0\label{6-13}\end{equation}
 with $\eta>0$, we obtain a lasing equation
 \begin{equation}
\frac{\partial}{\partial
t}\widetilde{d^{\dagger}}=\kappa^{\prime}\widetilde{d^{\dagger}}
-\eta\widetilde{d^{\dagger}}\widetilde{d^{\dagger}}\widetilde{d}.\label{6-14}
\end{equation}
Then the effect of the coherently injected electrons the SWCNT on
the light field is equivalent to that of a double-well potential
formed as
\begin{equation}
V(\left\vert d\right\vert )=-\kappa^{\prime}\left\vert d\right\vert
^{2}+\frac{\eta}{2}\left\vert d\right\vert
^{4},\label{6-15}
\end{equation}
Thus there exists a symmetry breaking based instability for laser
amplification. When $\kappa^{\prime}<0$, $d=0$ is the unique stable
point for the effective potential $V(\left\vert d\right\vert )$. In
this case we may safely neglect the nonlinearity, and the system is
only affected by stochastic processes. However, when
$\kappa^{\prime}$ passes through zero, the point $d=0$ is no longer
the stable point. Instead, the photon amplitude $d$ acquires its new
stable points with nonzero amplitude
\begin{equation}
\left\vert
d\right\vert=\sqrt{\frac{\kappa^{\prime}}{\eta}}
\end{equation}
indicating a phase transition in the system. The above phenomenon
that nonzero stable points of $V(\left\vert d\right\vert)$appear
means that a coherent light field with non-vanishing amplitude is
produced by the radiation of electrons confined in the SWCNT.

\section{\label{sec:conclusion}Conclusion}

In summary, our investigation in this paper is oriented by the needs
of designing the quantum devices in future. We theoretically studied
a solid state based quantum optical system, namely, the SWCNT
interacting with quantized light field. The ballistic transport of
electrons in SWCNT means quantum coherence of electrons in
terminology of quantum optics. Thus, the emitted and scattered light
from such coherent electrons could be quantum coherent as well, and
then we use the higher order coherence function to describe it. On
the other hand, SWCNT with different chirality $(n,m)$ have
different properties in their Rabi oscillations of the electrons
when driven by a strong single-mode light field. The anti-bunching
features of the light scattered by or emitted from them is also
studied in details here. The reason for such distinction of
chirality is that different sets of wave vectors $\mathbf{k}$ are
allowed in SWCNT with different chiral vectors, which may lead to
different energy structures in the SWCNT. Such effect is especially
significant on the $(2n,0)$ type SWCNT, where large degeneracy of
possible electron states onto $E=J$ occurs. This is a characteristic
property absent in 2D graphene. The possible lasing mechanism in the
SWCNT is also investigated theoretically, which may promise the
realization of nanoscale laser devices.

\appendix

\section{\label{App:strong}Semi-classical}

It is noticed that the semi-classical approximation applied in
Sec.~\ref{sec:Interband} is valid only for the quasi-classical case
in which the initial state possesses a very large number of single
frequency photons. We will justify this approximation with necessary
details in this appendix.

The complete dynamics of the SWCNT interacting with a strong light
field is displayed through the Schrodinger equations governed by the
Hamiltonian $H=H_{0}+H_{1},$ in the interaction picture, where \begin{subequations}
\begin{eqnarray}
H_{0} & = & \sum\limits _{\mathbf{k}}E_{\mathbf{k}}\left(\alpha_{\mathbf{k}}^{\dagger}\alpha_{\mathbf{k}}-\beta_{\mathbf{k}}^{\dagger}\beta_{\mathbf{k}}\right),\label{a1-2}\\
H_{1} & = & \sum_{\mathbf{k}}D_{\mathbf{k}}\left(de^{-i\Omega t}\alpha_{\mathbf{k}}^{\dagger}\beta_{\mathbf{k-q}}+h.c.\right)\label{a1-3}\end{eqnarray}
 And the initial condition of the system \end{subequations} \begin{equation}
\left\vert \Psi\left(0\right)\right\rangle =\left\vert \xi=\sqrt{N}e^{i\theta}\right\rangle \otimes\left\vert \phi\left(0\right)\right\rangle ,\label{a2}\end{equation}
 where the coherent state\begin{equation}
\left\vert \xi\right\rangle =\exp\left(\xi d^{\dagger}-\xi^{\ast}d\right)\left\vert 0\right\rangle \equiv D\left(\xi\right)\left\vert 0\right\rangle \label{a3}\end{equation}
 represents the state of the light field while $\left\vert \phi\left(0\right)\right\rangle $
stands for the initial state of the electrons in the SWCNT. We note
that $\left\vert \alpha\right\vert \simeq\sqrt{N}.$Since there is
no broken global phase symmetry, the arbitrary $\theta$ is chosen
as $0.$ The main reason for choosing the initial photon state as
a coherent one is that the average number $\left\langle \sqrt{N}\right\vert d^{\dagger}d\left\vert \sqrt{N}\right\rangle =N$
should be satisfied.

We introduce the photon vacuum picture, similar to the approach for
the semi-classical approximation of photon-atom system {[}cite
P.L.Kingt Concept of Quantum Optics], defined by \begin{equation}
\left\vert \Phi\left(t\right)\right\rangle
=D\left(\xi\right)^{-1}\left\vert \Psi\left(t\right)\right\rangle
,\left\vert \Phi\left(0\right)\right\rangle =\left\vert
0\right\rangle \otimes\left\vert \phi\left(0\right)\right\rangle
\label{a4}\end{equation}
 which satisfies the Schrodinger equation (in the interaction picture)
with the effective Hamiltonian\[
H_{e}=D\left(\xi\right)^{-1}HD\left(\xi\right)=H_{0}+V_{q}+H_{q}\]
 where \begin{subequations} \begin{eqnarray}
V_{q} & = & \sum_{\mathbf{k}}D_{\mathbf{k}}\left(\sqrt{N}e^{-i\Omega t}\alpha_{\mathbf{k}}^{\dagger}\beta_{\mathbf{k-q}}+h.c.\right)\label{a5-2}\\
H_{q} & = & \sum_{\mathbf{k}}D_{\mathbf{k}}\left(de^{-i\Omega t}\alpha_{\mathbf{k}}^{\dagger}\beta_{\mathbf{k-q}}+h.c.\right),\label{a5-6}\end{eqnarray}

Here $|0\rangle$ can be understood as a displaced vacuum. It should
be noticed that the above derivation is exact for the initial condition
(\ref{a4}).

For a very large $N$, $H_{q}$ in the above Hamiltonian is very small
with respect to the $V_{q}$, and it can be neglected in the first
order approximation. Under this approximation, the state of photons
is subjected to a collective evolution governed by the effective Hamiltonian
\end{subequations} \begin{subequations} \begin{eqnarray}
H_{e} & = & H_{0}+V_{q},\label{a6-1}\\
V_{q} & = & \sum_{\mathbf{k}}D_{\mathbf{k}}\left(\sqrt{N}e^{-i\Omega t}\alpha_{\mathbf{k}}^{\dagger}\beta_{\mathbf{k-q}}+h.c.\right).\label{a6-2}\end{eqnarray}
 Transforming back to the original picture, one proves the conclusion:
\textit{If} $N$\textit{ is a macroscopic number, namely, it is large
enough, the total system will evolve with a factorizable wave function}
$|\Psi(t)\rangle=|\sqrt{N}e^{i\theta}\rangle\otimes|\phi(t)\rangle,$
\textit{where} $|\phi(t)\rangle$ o\textit{beys the Schrödinger equation
governed by the effective Hamiltonian} $H_{e}$.

The next question is the effects of the neglected term, $H_{q}$,
on the dynamics in the photons vacuum picture. In the framework of
the perturbation theory, the role of $H_{q}$ relies on the coupling
to the vacuum, that is \end{subequations} \begin{eqnarray}
H_{q}|\Phi(0)\rangle & = & \sum_{\mathbf{k}}D_{\mathbf{k}}\left(de^{-i\Omega t}\alpha_{\mathbf{k}}^{\dagger}\beta_{\mathbf{k-q}}+h.c.\right)|0\rangle\otimes|\phi(0)\rangle\notag\\
 & = & e^{i\Omega t}|1\rangle\otimes\sum_{\mathbf{k}}D_{\mathbf{k}}\beta_{\mathbf{k-q}}^{\dagger}\alpha_{\mathbf{k}}|\phi(0)\rangle\label{a7}\end{eqnarray}
 which leads to a single-particle excitation of the vacuum. Finally
we reach the following conclusions: (1) In the large $N$ limit, this
excitation is weak compared with the collective motion; (2) If there
is initially no collective excitation or single excited electrons
in the SWCNT, the system will be stable and remain in the displaced
vacuum state even when $H_{q}$ is taken into account.

\begin{acknowledgements} The authors thank H. Dong for his schematic
diagram of the graphene. This work is supported by NSFC No.10474104,
No.60433050, and No.10704023, NFRPC No.2006CB921205 and
2005CB724508.
\end{acknowledgements}

\end{document}